\documentclass{article}
\usepackage{graphicx}
\usepackage{graphics}
\usepackage{amsfonts}
\usepackage{marvosym}

\begin{document}
\title{A History of Cluster Analysis Using the 
Classification Society's Bibliography Over Four Decades}
\author{Fionn Murtagh (1) and Michael J.\ Kurtz (2) \\
(1) School of Computer Science and Informatics \\
De Montfort University \\
Leicester LE1 9BH, UK \\
(2) Harvard-Smithsonian Center for Astrophysics, \\ 60 Garden Street,
Cambridge, MA 02138, USA \\
Email: fmurtagh@acm.org}

\maketitle

\begin{abstract}
The Classification Literature Automated Search Service, an
annual bibliography based on citation of one or more of a 
set of around 80 book or journal publications, ran from 
1972 to 2012.  We analyze here the years 1994 to 2011.   
The Classification Society's 
{\em Service}, as it was termed, has been produced by the 
Classification Society.  In earlier decades it was distributed
as a diskette or CD with the {\em Journal of Classification}.
Among our findings are the following: an enormous increase in 
scholarly production post approximately 2000; a very major 
increase in quantity, coupled with work in different disciplines, 
from approximately 2004; and 
a major shift also from cluster analysis in earlier times having
mathematics and psychology as disciplines of the journals published
in, and affiliations of authors, 
contrasted with, in 
more recent times, a ``centre of gravity'' in management and engineering. 
\end{abstract}

\section{Introduction}

Clustering as a problem and as a practice in many different domains 
has proven to be quite perennial.  Testifying to this is the 
presence of ``clustering'' or ``cluster analysis'' as a term in 
an important classification system. The premier professional 
organisation in computing research, the ACM (Association for 
Computing Machinery), has a standard classification labelling 
system for publications.  Released in 1998, a major update was
released in Sepember 2012, and another release is expected in 2014.
In the 2012 ACM Computing Classification System (CCS, 2012), part of 
the category tree, at increasing level of detail, is as follows: 
``Mathematics of Computing'', 
``Probability and Statistics'', ``Statistical Paradigms'', ``Cluster 
Analysis''.   The 1998 Computing Classification System (CCS, 1998)
had clustering included in category H.3.3, and I.5.3 was another 
category ``Clustering''.   

\begin{figure}
\begin{center}
\includegraphics[width=12cm]{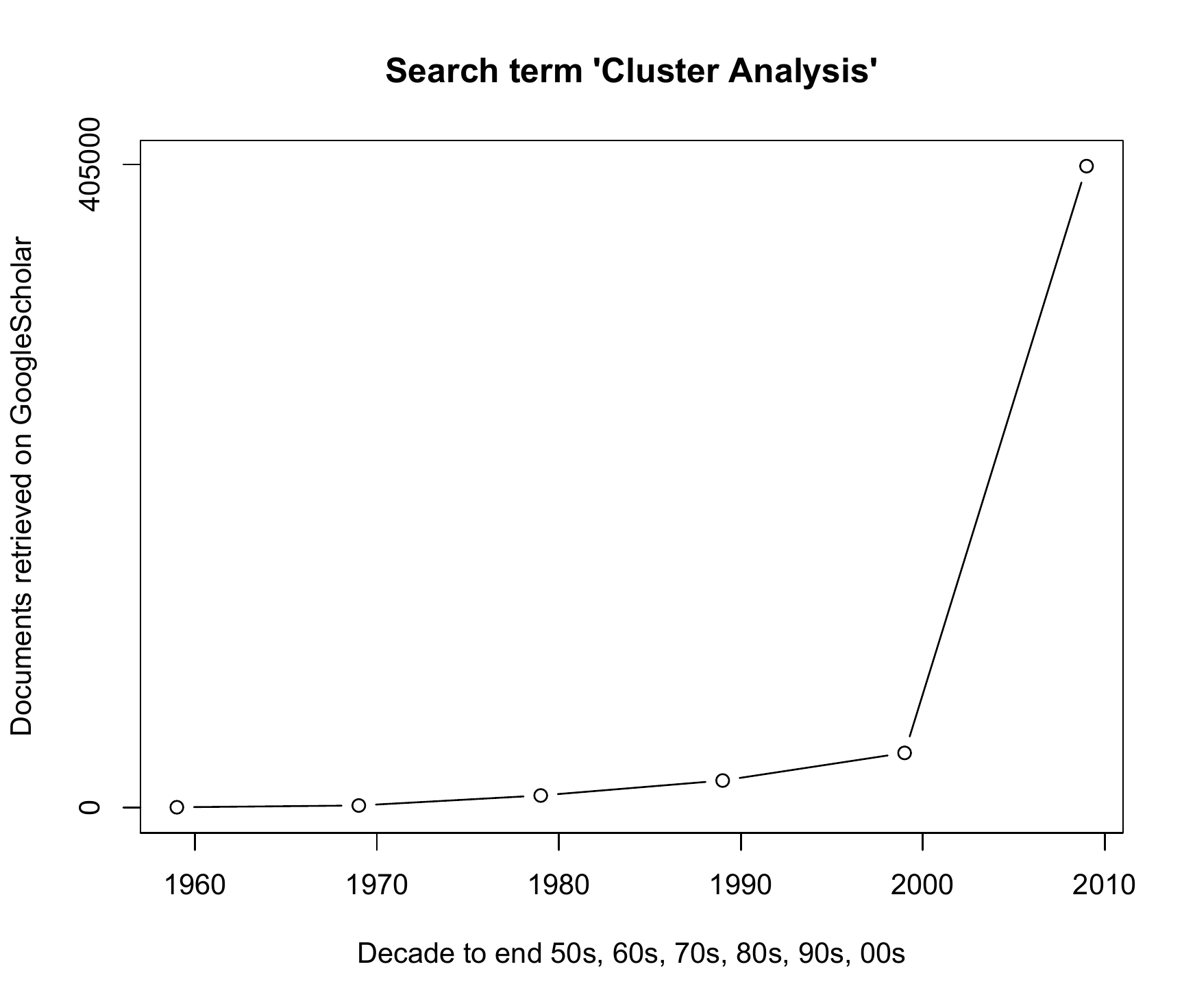}
\end{center}
\caption{Google Scholar retrievals using search term ``cluster
analysis'', for the years 1950-1959, 1960-1960, etc., up to 2000-2009.
(Data collected in September 2012.)}
\label{figgs1}
\end{figure}

\begin{figure}
\begin{center}
\includegraphics[width=12cm]{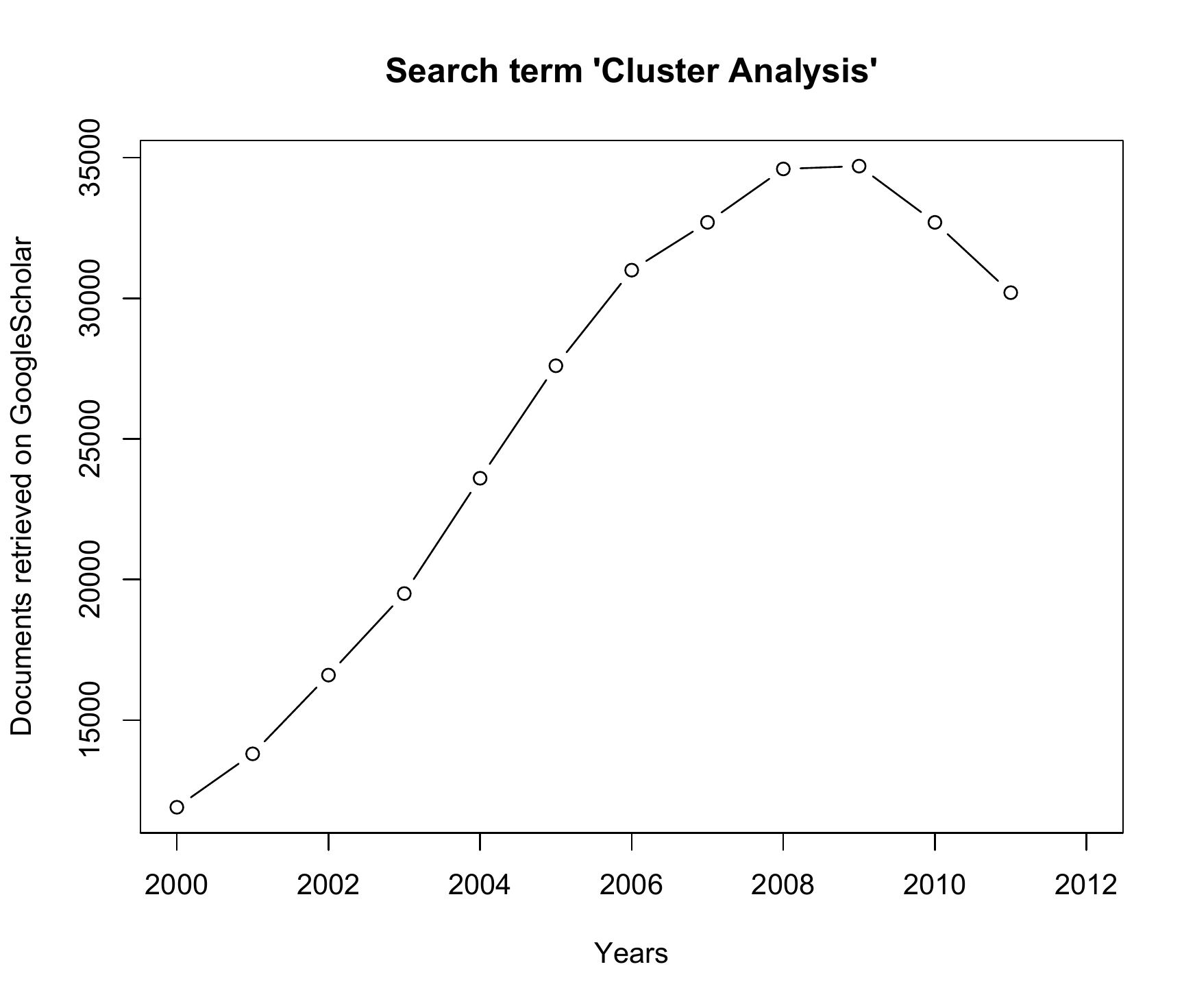}
\end{center}
\caption{Google Scholar retrievals using search term ``cluster
analysis'', for the years 2000, 2001, 2002, $\dots$, 2011.
(Data collected in September 2012.)}
\label{figgs2}
\end{figure}

Figures \ref{figgs1} and \ref{figgs2}, using the Google Scholar 
content-searchable holdings,  present a view of this 
perennial and mostly ever growing use of clustering.  The term
``cluster analysis'' was used.  Documents retrieved, that use that 
term in the title or body, increased to 404,000 in the decade 2000-2009.  
 Of course lots of other 
closely related terms, or more specific terms, could additionally 
be availed of.   
These figures present no more than an expression of the growth 
of the field of cluster analysis.   The tremendous growth in 
activity post-2000 is looked at in more detail in Figure 
\ref{figgs2}.  Time will tell if there is a decrease in use 
of the term ``cluster analysis''.   Again we note that this 
is just one term and many other related terms are relevant too.   

A sampling of historical overviews of clustering follow.  
Kurtz (1983) presents an overview for the astronomer 
and space scientist.  The orientation towards
computer science is strong in Murtagh (2008), which takes 
in linkages to the Benz\'ecri school of data analysis, and also 
current developments that have led Google's Peter Norvig to claim
(with some justification, albeit very debatable) that similarity-based
clustering has led to a correlation basis coming to the fore
in science, potentially replacing entirely the causation principle.   
A general history is also presented in Murtagh (2013).

\section{40 Years of Service, the Classification Literature 
Automated Search Service}
 
The Classification Literature Automated Search Service
started with Volume 1 in 1972 (using the previous calendar
year's data).    
It was published as a printed booklet from the start 
and bore the ISSN 0731-4043.  It was necessary 
in the editorial and printing team to plan carefully the
total number of printed pages, to liaise with Springer's 
headquarters and distribution center 
(these were in New York City and in Secaucus, New Jersey)
 and the printing company 
used (Sheridan Press, Pennsylvania) and then to have copies 
of the bibliography shipped 
with the {\em Journal of Classification} to subscribers and 
also to libraries. 
 
From June 1984 through May 1993,   
the Editor of {\em Service}, or sometimes {\em CSNA Service} as 
it was referred to, was William H.E. Day.  During that time 
the Technical Editors were Elaine Boone (1984--1988) and Eva
Whitmore (1989--1993).  Bill Day also had the assistance of 
Todd Wareham, a computer science student then, in the preparation
of camera-ready copy for volumes 14--16 (1985--1987).  
Technical Support included use of C programs and Unix scripts for 
reformatting the data.  
Bill Day
was based in the Department of Computer Science, Memorial 
University of Newfoundland, St.\ John's, Newfoundland.
%where Todd is a Professor 
%now, and Eva was based in St.\ John's too in those earlier years.  
% MSG FROM TODD, 23AUG2012:
% please refer to me as ``Todd Wareham'' in the article 
%though ``H. Todd'' is technically correct, I go by ``Todd'' in everyday and
%professional life,

The data was obtained from ISI, 
the Institute for Scientific Information, which is now a subsidiary 
of Thomson-Reuters (and publishes the World of Science, the Science 
Citation Index, the Social Science Citation Index, and other products). 
Processing of the bibliographic data was always required.  In the early 
years, a range of nroff and troff text processing utilities were used to 
re-format the data.   Due to limitations 
on the output and distributed format (book, later diskette, then CD), various
algorithms were applied to restrict the quantity of data.  This included 
filtering by listing journal titles, and keywords to be 
excluded in titles of published articles.  Thus, in the latter case,
medical terms, or 
``galaxy cluster'', betokened non-algorithmic matters and hence were to be
excluded.

Before Bill Day, Roger K. Blashfield (University of Florida) was Editor.  
% Bill: I have no other information concerning the publication of 
%volumes 1-11.
Fionn Murtagh was Editor from 1993 to 2008.  Michael Kurtz was Editor 
thereafter.  Eva Whitmore remained as Technical Editor, having started 
as noted above in 1988.  

In the 1990s the bibliography went to diskette 
format and there too we rapidly went to the storage capacity of 
the media at that time -- 5.25 inch ``floppies'' that were to be 
replaced by 3.5 inch diskettes.  It made sense then, as announced
by us in the production team in October 
1999, to transit to CDs, which additionally allowed us, due to the
storage available, to have previous years' bibliographies, and then 
to have scanned copies of ``profile'' books available on the CD.  Below
it is explained just how the profile of books and articles was used to 
drive the retrieval process and thereby to define the domain of 
interest.  

In 1994, on-line content search to the bibliographies was supported 
by the WAIS, Wide-Area
Information System,  distributed search and retrieval standard.
This was an early forerunner of the search engines to come a 
few years later.
% See http://www.pitt.edu/~csna/wais.html

As either book
or as CD, the bibliography, that was termed {\em Service}, 
was distributed as a supplement to the 
{\em Journal of Classification}, published by Springer
on behalf of what is now called the Classification 
Society\footnote{http://www.classification-society.org/clsoc}.  The 
Classification 
Society was set up in April 1964.   In December 1968, European 
and North American branches were set up, and were largely autonomous.  
These branches became the Classification Society of North America
(CSNA), 
and the British Classification Society. In 2008, CSNA reverted to 
its former name of the Classification Society.  (Various historical
documents can be found on the web sites of the Classification 
Society, and of the British Classification Society\footnote{Currently
http://thames.cs.rhul.ac.uk/bcs}.)  
 The 
{\em Journal of 
Classification} saw its Volume 1, Number 1, in 1984, 
and its first Editor-in-Chief was Phipps Arabie (born 
13 March 1948, died 23 June 2011).  Phipps Arabie was a very strong 
supporter always of the {\em Service} bibliographies. From 2002 to date
(2012), Willem Heiser is Editor-in-Chief of the {\em Journal of 
Classification}.

In 2008, the last CD was produced, due to the plan 
be be web-based only.  In 2012, ISI 
discontinued the provision of data completely.  
For online access now to {\em Service}, see 
{\tt https://www.cfa.harvard.edu/$\sim$kurtz}

\begin{figure}
\begin{center}
\includegraphics[width=4cm]{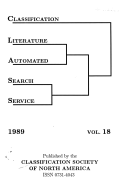}
\end{center}
\caption{A typical cover of the printed volume of {\em Service}.}
\label{figcover}
\end{figure}

The following\footnote{Currently 
available as file service23.profile.txt at address
ftp://ftp.pitt.edu/group/csna.} 
is from the introduction to Volume 23, 1994.  It explains the mechanism 
used to carry out the searches and to assemble the bibliographies.

\begin{quote}
``This volume of the Classification Literature Automated Search Service 
contains a bibliography of 2497 classification-related journal papers 
which appeared in 1994.  In order to use the Service knowledgeably for 
reference, readers should know about the databases from which the journal
papers were selected, the criteria employed to identify 
classification-related papers, and the mechanisms provided to access 
bibliographic information about classification-related papers.

The first step in constructing the bibliography is to collect data about 
journal papers. The Service obtains these data from Research Alert, a 
bibliographic service of the Institute for Scientific Information (ISI, 
Philadelphia).  Research Alert enables the Service to access papers in 
over 8000 science, technical, and social sciences journals including 
those from the Science Citation Index, Social Sciences Citation Index, 
and Arts  Humanities Citation Index databases.  To use Research Alert, 
the Service's Editorial Board compiles a list, or profile, of books or 
papers, called profile items, that are indicative contributions to the 
theory and practice of classification.  A journal paper is considered to
be classification-related if it cites one or more profile items.  Using 
the profile, Research Alert provides the Service with bibliographic 
information about classification-related papers in recent issues of the 
8000 journals it processes.
   
Research Alert's selection procedure depends completely on the profile.  
The Editorial Board reviews the profile regularly in order to ensure
that the papers selected by Research Alert are relevant to classification 
and related areas of data analysis.  The profile for this volume of the
Service contained 82 items and appears in the file `profile.txt'.  The 
Editor welcomes your suggestions for improving the composition of future
profiles.''
\end{quote}

\section{The Data: The Bibliographies of Clustering from 
1994 to 2011} 

The bibliography was always collected for the previous year.  Hence 
volume 23, corresponding to 1994 data, was published in 1995.  The 
volumes continue up to volume 40, with 2011 data, consolidated in 2012.  
In volumes 23 to 40, there are 85,020 bibliographic records.  
% /Users/fionnmurtagh/MB-idir-laamha/CRC/MichaelKurtz/PREPARE-XML-FILE
% grep biblio biblio*-Solr.xml | wc  BUT THEN HAD TO SEE THAT 16 WERE LEGIT
% uses of ``biblio'' in title or keywords. 

\subsection{The Profile Publications Used to Drive
the Search and Retrieval}

The ``profile'' publications used in the past few years are
listed in the Appendix.   Citing any one of these publications was
therefore the criterion used for assembling the annual bibliography.    

The number of citations per year is shown in Table 
\ref{tab1}.  This relates to the bibliographies for the years
1994 to 2011.   

It is to be noted how some of the profile 
publications were introduced in a given year.  (Consider, 
for example, Blashfield76, introduced from 2004.)  

In the 
case of, for example, Bishop95, in pre-1995 years, 
the search term used here
(``BISHOP CM'') picked him up as an author of another publication 
and not the profile publication, his 1995 book.  
%Thus in the data used
%here there will be some imprecision as to what the profile publications
%are picking up, but it will be not only small scale, but also one 
%might expect that as a regular author the presence of Chris Bishop,
%as just discussed, should be counted.  

Note too that the (different) 
works of some authors are combined by us.  Such is the case for example for 
two publications by Doug Carroll, published in 1970 and in 1980.  (J.\ Douglas
Carroll, 1939--2011, worked most recently at Rutgers Business School.
He was an early developer of, and founder of the field of, multidimensional
scaling and other methods and their applications in psychometrics.)

%Table \ref{tab2} was derived from Table \ref{tab1} by removing any profile 
%item that came into 
%use late relative to 1994, or was removed from some year onwards.  
%Table \ref{tab2} provides a complete data set for us to look at in 
%more detail.  

\begin{table}
\tiny{
\begin{verbatim}
                   94  95  96  97  98  99  00  01  02  03  04   05   06   07   08   09   10   11
Adams72             8  12   7   7   5   4   2   5   5   5   5    2    5    6    3    5    3    3
Anderberg73        73  54  67  60  64  78  72  58  65  56  39   56   48   64   53   77   52   67
Arabie87            8   7  11   8  10  10   5   4   6   4   0    0    0    0    0    0    0    0
Avise74            11  12  15   9   9  10   5   7   1   3   1    6    2    1    3    0    4    3
Benzecri73         41  50  50  40  51  43  51  30  29  39  24   30   32   25   32   34   38   35
Bezdek81           45  45 107  87 114 131  97 118 126 151 337  432  496  495  558  701  588  551
Bishop95            2   0   1   0   1   1   0  10 262 335   0    0    0    0    0    0    0    0
Blashfield76        0   0   0   0   0   0   0   0   0   0   3    6    8    4    3    8    5    5
Breiman84         106 130 181 165 199 228 224 224 275 358 266  294  345  393  441  620  528  647
Carroll70,80       34  24  25  31  24  40  27  40  33  28   0    0    0    0    0    0    0    0
Cormack71           2   6   7   2   4   7   5   4   8   2   7    8    3    5    8    9    6    9
Cover67            24  21  21  24  28  30  28  31  25  54   0    0    0    0    0    0    0    0
Devijver82         32  28  42  59  54  37  32  43  31  47  64   63   72   50   64   62   43   49
Diggle83           33  35  41  36  38  49  57  51  56  56  77   87   68   78   84  107   97   73
Duda73            241 252 304 283 337 309 289 324 379 600 825 1033 1175 1309 1416 1554 1164 1316
Efron83            14  24  21  21  20  30  32  26  16  25  20   43   60   61   56   66   56   53
Eldredge80         25  21  24  24  25  19  15  16  17  23  38   30   19   32   43   31   37   38
Everitt79,80       33  52  47  35  38  39  37  41  26  18 132  146  139  133  148  181  186  166
Farris72           22  23  39  26  21  25  12  21  18   9   5   10    4   12    9   10    3    7
Felsenstein82      35  38  23  16  16  13   6  16   7  10   3   10    3    5    9   10    4    3
Fisher36           42  49  50  70  64  68  60  67  86  96  96  207  187  223  249  255  267  284
Fitch67           102 104 113 111  90  79 110  75  85  65  63   84   64   51   62   47   44   36
Friedman77         10   8  13  13  18  18  18  12  17  17  26   28   23   30   40   38   38   34
Fu74,82            33  26  27  20  29  32  22  23  10  23  58   58   58   52   57   58   46   58
Fukunaga72         28  38  43  38  37  34  29  31  36  34 374  447  396  478  528  549  424  455
Gauch82            73  80  80 104  76 104  77  91  90  88   0    0    0    0    0    0    0    0
Gnanadesikan77     15  26  13  15   8  17  19  12  14  11   0    0    0    0    0    0    0    0
Gordon81           21  23  18  16  20  12  20  16  16   8  65   73   66   70   61   87   56   60
Gower66            40  14  33  37  27  34  34  33  33  47  29   52   66   60   60   87   61   95
Greenacre84        46  56  69  50  69  71  51  52  49  61  25   41   34   28   33   36   40   39
Guttman68          25  16  11  14  18  13  14  15  12  16  14   14   20   16   20   20    9    7
Hand81             33  26  30  28  30  33  20  18  28  25  65   89   96   83   95  119  102  119
Hartigan75         68  64  76  62  62  72  79  87  86  94 143  150  145  160  195  244  178  203
Hennig66           86  79  87  93  80  76  81  72  72  79 102  100  105  103  115  110   97  130
Hill74             14  15  19  14  13  10  11  11   4  12   0    5    9    3   18   14   11    2
Huber85            22  24  29  18  19  37  18  27  20  24   0    0    0    0    0    0    0    0
Hubert7685         15  12  14  16  13  19   8  23  18  37  36   48   72   78  104  135  137  134
Jain88             52  44  54  58  59  81  74  81  81 115 716  860  910  939 1014 1190  882  969
Jardine71           6   6   6   9   7   4   5   8   6   6  14   21   13   18   25   32   24   26
Johnson67          24  27  20  18  10  13   7   7   9   7  11   29   28   37   55   48   40   48
Kluge69            58  42  55  56  66  63  79  96  58  85  59   78   81   85   99  115  105  112
Kohonen95           0   0   2   0  13  52  88  80  91  97 374  444  415  450  534  693  438  462
Kruskal64,78       72  78  90 102  99  98 112 119 123 141 152  186  204  207  255  255  223  278
Lance67            11  13  14  15  10  22   8  18  13  16  18   24   17   13   21   17   15   25
Legendre83         20  24  33  31  32  25  36  24  22  21  26   44   33   42   45   24   35   37
Lorr83             11  19  13   6  10   8   5   7   9  11  11   19   12   19   12   13   10   17
Maddison84         55  48  41  38  48  50  46  34  35  23   0    0    0    0    0    0    0    0
Mantel67           61  77  99 117 112 138 145 185 182 207 163  296  308  360  362  453  392  452
Mayr69             26  22  19  31  22  25  26  21  19  15 132  140  124  152  142  149  128  179
McLachlan88,92,97  28  33  37  33  33  28  34  50 130 136 154  137  168  188  173  213  234  235
Michalski83        53  45  52  38  43  38  31  21   9  15  48   30   28   22   31   45   33   26
Milligan80,81,85   39  35  35  40  34  36  29  39  66  63  56   58   84   79   91  105   88  106
Murtagh83           2   2   0   0   4   4   0   2   1   0   6    8    8    8    9   10   15   15
Nei72             139 140 170 179 156 169 185 172 169 166 131  188  185  193  198  198  191  220
Nelson81           45  32  46  27  38  33  34  39  49  30  46   61   60   50   60   77   52   62
Nosofsky84         14  15  14  16  25  10  15  11  21  20   8   16   17   14   17   10   23   27
Orloci78           14   7  15  15  12  10  14  10   6   9   8   10   18   17   11   22   25   26
Pavlidis77         14  15  18   9  12  21   7   8  10   6  90  103   82   59   90   91   52   60
Punj83             11  13  14  11  13  12   9   8  15  20  17   23   21   18   24   51   30   45
Rammal86           13   9   9   7  16  11  12  14  14   5   0    0    0    0    0    0    0    0
Rand71              4   8   4   6   2   3   2  11   9  11  17   33   39   51   78  118  104   85
Reyment84          13  15  10  17  12  12  10   4  10  10  12   18   14   21   15   17   10   18
Ripley81           50  46  52  63  59  64  60  63  60  57 173  192  193  175  179  244  178  215
Rohlf82            10   6   7  10   3   3   6   4   4   6   2    2    2    7    2    4    2    3
Sammon69           22  27  29  36  32  47  45  47  43  63   0    0    0    0    0    0    0    0
Sankoff83          31  24  31  27  38  28  36  27  28  26  55   68   76   72   64   58   71   59
Sattath77           8  10  15   9  12   8  12  16   9  11   7    9    8    9   11   11    5    9
Schiffman81        22  32  34  33  25  21  20  28  19  24  20   17   14   12   18   18   14   17
Silverman86       117 131 183 140 161 192 172 158 191 206 264  306  287  350  384  462  410  383
Sneath73          385 374 422 435 406 386 367 360 355 357  94   99  111  139  122  133   92  128
Sokal63            53  52  45  53  42  35  44  43  40  36 225  272  251  252  287  282  225  282
Spaeth80            9  13  10  17  10   6  10  12   9  16  17   32   26   24   29   43   35   32
Spitzer74          12   5   5   2   4   2   2   4   3   8   6    3    3    7    4    9    5    4
Swofford81        116 127 171 151 135 127 125 115 102  84  55   67   46   36   32   20    9   24
Tversky77          46  68  65  82  69  81  74  70  63  97  62   97  109  101  126  142  134  136
VanLaarhoven87     46  56  65  45  45  42  48  44  33  32  54   65   58   64   53   64   38   61
VanRijsbergen79     0   0   2   0   0   0  46  54  38  88   2    2    8    7    9   19   12   18
Ward63             65  74  75  81  81  95 109 115 100 130 109  191  182  223  231  303  274  318
Wiley81            79  66  52  66  61  54  59  53  43  60  47   71   53   69   75   66   56   68
Wishart87          29  16  12  18  10   8  17   7  14  14   7    9   15   17   16   19   23   12
Wolfe70             6   9   8   8   8  11   8   8   9   4   7    7    8    7    5   13    8   16
Zahn71              8  11  14  13  11   5  10   8   9  21  20   24   14   22   24   29   30   25

\end{verbatim}
}
\caption{Frequencies of occurrence found for the 82 profile publications, 
over 18 years.  There are 135,088 bibliographic citations in all, i.e.\ 
the grand total of this table.}
\label{tab1}
\end{table}

\subsection{Changed Data Provision After 2003}

%VanRijsbergen79 was VANRIJSB.CJ, VANRIJSBERG and RIJSBERGEN CJV in succession.
%Others (Silverman86, Schiffman81) dropped second initials.  VanLaarhoven87 
%dropped a last letter in that name.  

\begin{figure}
\begin{center}
\includegraphics[width=10cm]{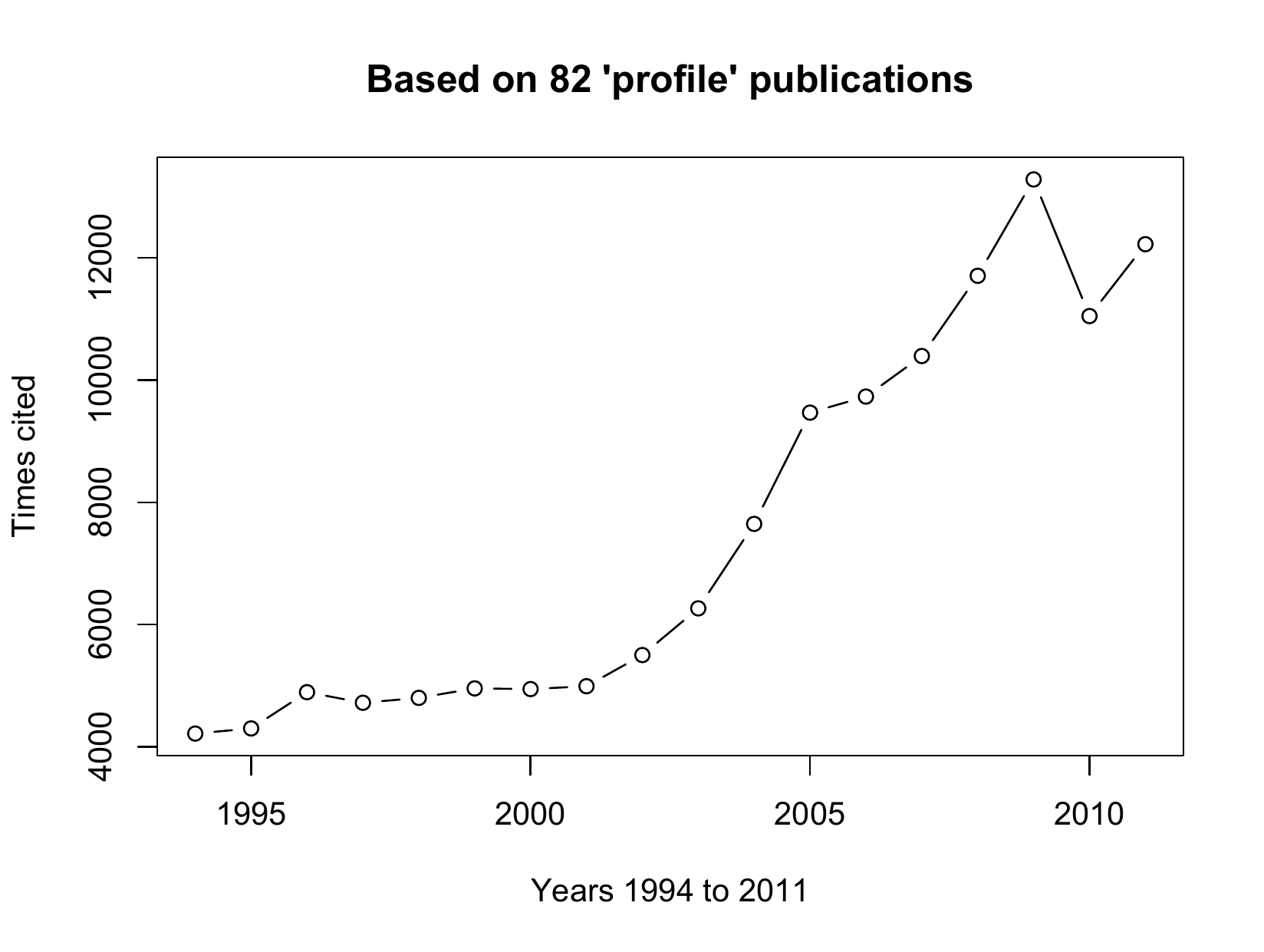}
\end{center}
\caption{Increase of citing bibliographic entries over the years.}
\label{fig1}
\end{figure}

What is particularly noticeable about Table \ref{tab1} is the increase in 
citations over time.  See Figure \ref{fig1}. 
While it is the case that (i) there was some net increase each 
year, but (ii) nonetheless the lack of
constraint related to distribution medium from 2004, 
(iii) that can be coupled 
with the massively 
growing volume of research production worldwide, and finally (iv) 
the high point of 2009, 
maybe given a lag to be expected in publishing following the economic 
downturn in the Western countries that  
started in 2008.  
Among other changes from 2004, see how Sneath73 is replaced largely (cf.\
Table \ref{tab1}) by the earlier jointly authored volume, Sokal63.   
(P.H.A.\ Sneath,
University of Leicester, UK, 17 November 1923 -- 9 September 2011, and 
 R.R.\
Sokal, State University of New York, 13 January 1926 -- 9 April 2012, 
were key names in the development of numerical taxonomy.) 
%See \cite{hull}
%for a wide-ranging intellectual history, including the roles played 
%by Sneath, Sokal, F.J. Rohlf and other key figures in cluster analysis.  

Table \ref{tab2} is an alternative view of the 18 years we are dealing
with.  This table shows a range of discipline names that are picked up 
in the data by their appearance in a journal title, or a publication 
title, or an area title.   The terms used are: Medicine, Biology,
Physics, Chemistry, Astronomy, Mathematics, Statistics, Engineering, 
Psychology, Psychiatry, Literature, Humanities, Economics and Sociology.

\begin{table}
\small{
\begin{verbatim}
       94  95  96  97  98  99  00  01  02  03  04  05  06  07  08  09  10  11
Med     1   0   1   2   2   1   3   3   5   5  82 111 120 136 147 205 146 156
Bio   334 345 372 369 389 364 387 346 397 392 274 354 360 393 463 559 508 545
Phys    5   5   1   2   4   1   4   5   9   5  73  60 104  94 129 113 132 131
Chem   76  55  64  59  88  63  88  87 100 117  57  83  86  71  93 138 125 115
Astr    1   0   1   0   1   0   0   2   3   1   4   4   9   4  14   5   3  19
Math   70  63  84  63  82  89  77  79  86  79  23  42  27  39  41  55  48  62
Stat   98 105 117  94  92 128 111 104 125 101 207 255 266 287 312 340 345 310
Eng   108 118 121 137 135 132 134 161 216 217 355 424 468 470 753 517 472 689
Psych 128 155 164 141 128 132 147 118 124 151  77  72  85  94  97 103  93  99
Psy     2   0   0   1   0   0   0   2   1   0  15  12   6  12  10  13  10   9
Lit     2   1   1   1   3   1   3   2   1   5   9   9   8   4   3   7   9  10
Hum     9  19  15  15  18  14  23  17  19  36   1   0   1   1   4   4   7   4
Eco     0   0   0   0   2   0   0   2   1   1  26  29  33  30  40  53  45  42
Soc    21  24  27  16  25  17  17  24  24  20   7   2   5  16   2  13  11  12
\end{verbatim}
}
\caption{Frequencies of occurrence found for the discipline terms -- used
in journal or article titles or otherwise.  The columns are the years 
1994 to 2011.  See text for the spelling out of the disciple labels.  The 
number of occurrences here is, in all, 23,997.} 
\label{tab2}
\end{table}

From the editorial report of {\em Service} to the Board 
of the Classification Society of North America in June 2003,
there is the following explanation of the sea-change in the 
bibliographic source data from 2004 onwards.  

\begin{quote}
``The CD containing Service data, cumulative over a number of years,
with a Java graphical user interface, and copies of Hartigan's
(scanned) and van Rijsbergen's classical books, was distributed 
as usual with issue 1 of the Journal of Classification.  Number of 
copies produced 525.  ...

Up to now, Eva got the data quarterly and initially processed it at 
Memorial University.  (A long time ago Bill Day there was the link with 
Memorial).  Probably the scripts in use there are 15 years old, or
more.  Then I did some processing, with a number of Unix scripts.  
For the CD, a Java application based search GUI was written 
2 or 3 years ago, and of course assumed the particular format 
discussed above.  Now ISI, from whom we purchase the data (about 
USD 70 per profile item) are changing the dissemination mechanism 
and the format. ...

... our new format for receiving data from ISI.  
[ISI] emailed me about a week ago and informed me that 
``Research Alert'' data will no longer be available -- they are 
switching totally to ``Personal Alert'', as below. We get the same 
data, in a weekly email, but as you can see, the format is         
different.                                         

[...]

I notice this data has keywords associated.'' 
\end{quote}

Tables \ref{tab5} and \ref{tab6} are indicative of these formats.

\begin{table}
\begin{verbatim}
T       Learning to Set-Up Numerical Optimizations of
T       Engineering Designs
A       SCHWABAC.M
A       ELLMAN T
A       HIRSH H
K       MATHEMATICAL SCIENCES - Computer Science
U       AI EDAM      12(2): 173-192,APR 1998
W         M Schwabacher, Natl Inst Stand &
W         Technol, Gaithersburg, MD 20899
W.      BREIMAN L    84
\end{verbatim}
\caption{``Research Alert'' format for the bibliography data used up to 2003.
First entry of Volume 27, 1998.  AI EDAM is the Cambridge University Press
{\em Artificial Intelligence for Engineering Design, Analysis and 
Manufacturing}.  The ``profile'' entry is Breiman, Friedman, Olshen 
and Stone, {\em Classification and Regression Trees}, Wadsworth, 1984.}
\label{tab5}
\end{table}

\begin{table}
\begin{verbatim}
TITLE:          Multiscale spatial variation of the bark beetle Ips
                sexdentatus damage in a pine plantation forest (Landes de Gascogne,
                Southwestern France) (Article, English)
AUTHOR:         Rossi, JP; Samalens, JC; Guyon, D; van Halder, I;
                Jactel, H; Menassieu, P; Piou, D
SOURCE:         FOREST ECOLOGY AND MANAGEMENT 257 (7). MAR 22 2009.
                p.1551-1557 ELSEVIER SCIENCE BV, AMSTERDAM

SEARCH TERM(S):  RIPLEY BD  rauth; DENSITY ESTIM*  rwork; MULTI*  rwork

KEYWORDS:       Bark beetle; Ips sexdentatus; Pinus pinaster; Spatial
                statistics; Ripley's statistic; Aggregation; Landscape;
                Plantation forest
KEYWORDS+:       POINT PATTERN-ANALYSIS; TYPOGRAPHUS L.; FELLED TREES;
                SPRUCE; SCOLYTIDAE; COLEOPTERA; MANAGEMENT; WINDTHROW;
                RISK; COLONIZATION

AUTHOR ADDRESS: JP Rossi, INRA, UMR BIOGECO, Domaine Hermitage 69 Route
                Arcachon, F-33612 Cestas, France
\end{verbatim}
\caption{``Personal Alert'' format for the bibliography data used from 2004.
First entry of Volume 38, 2009.  The ``profile'' item is B.D. Ripley,
{\em Spatial Statistics}, Wiley, 1981.}
\label{tab6}
\end{table}

\section{Semantic Analysis of Profile Publications and of Disciplines, 
over 18 Years}

\subsection{Semantic Analysis}
\label{semantic}

Take the observables, e.g.\ profile publications, or disciplines,
 as indexed by $i$.  Take the attributes, e.g.\ the years, as indexed by 
$j$.   Call the mass of observable $i$ to be 
$f_i$, and analogously the mass of 
attribute $j$, $f_j$.  These masses are components of marginal distributions.
Alternatively expressed, the $f_i$ and $f_j$ terms, for all $i$ and $j$,
are respectively 
the empirical probability distribution defined on the set of all 
observables, $i$, and on the set of all attributes, $j$.   The domains
of the function $f$ are thus, 
respectively, the observables set and the attributes set.  

The frequency of occurrence data used for observable $i$ and attribute $j$
is $f_{ij}$.  Correspondence Analysis is firstly and foremostly the 
study of discrepancy of $f_{ij}$ from a sort of null hypothesis expressed by 
$f_i f_j$.   

A successively best fit Euclidean representation is found, to embed the 
observable set, and the attribute set.  Let the observable $i$ have 
embedding, firstly, and then, secondly, projection $\psi_i$ relative
to factor $\psi$, and similarly for attribute $j$ relative to factor $\phi$. 
The associated eigenvalue of the pair of factors $\psi_i$ and $\phi$ is
$\lambda$.  

We require the semantic relationship tying together observables and
attributes vis-\`a-vis each successive factor:

$$ \sqrt{\lambda} \psi_i = \sum_j {f_{ij} \over f_i} \phi_j 
\mbox{ \ \ \ \    and in the dual space  \ \ \ \   }
 \sqrt{\lambda} \phi_j = \sum_i {f_{ij} \over f_j} \psi_i $$
These are termed {\em transition formulas}.   

Supplementary elements, rows or columns, 
 are when we use $f_{ij}$ values that are, through these relationships,
 projected post hoc into the analysis.  

The semantic analysis framework is now used to provide (1) visualization,
seeking particular salient interrelationships in the data, and (2) 
summarization of the data through clusters, where we use years, 
disciplines and publications to achieve a good understanding of the data.
Here (1) is a planar, and hence low-dimensional, expression of the 
data, whereas in (2) the clustering is carried out in the data's
full dimensionality.  

\subsection{Major Change: Pre-2004 and From 2004 Onwards}

\begin{figure}
\begin{center}
\includegraphics[width=12cm]{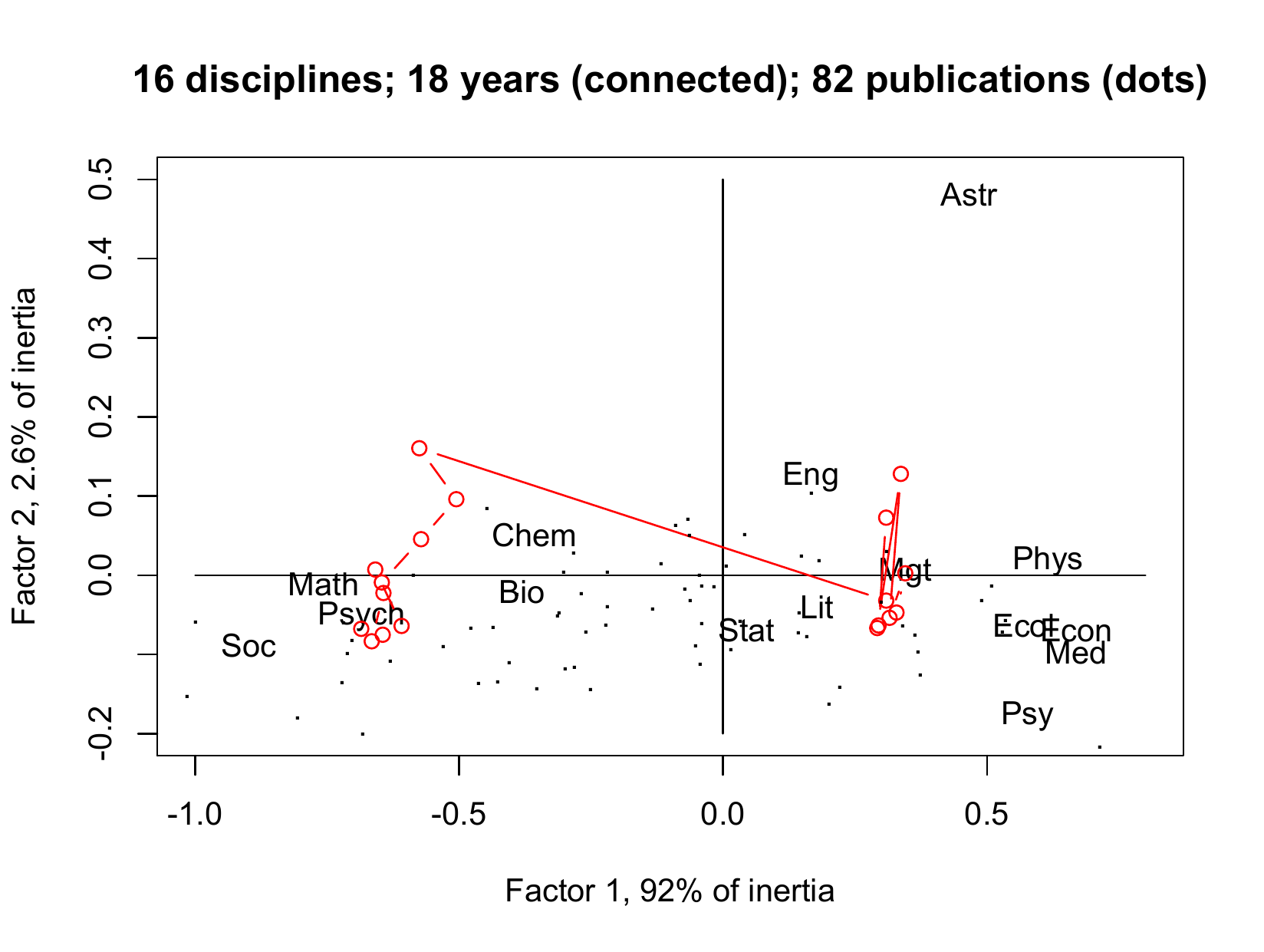}
\end{center}
\caption{Principal factor plane of a Correspondence Analysis of 
(primary analysis) 16 disciplines crossed by 18 years; and projected
into this (supplementary elements) 82 ``profile'' publications.
(Medicine is very slightly repositioned to avoid overlap.
Humanities is off the figure to the left, at coordinates $-1.168, 
0.251$.)}
\label{figCA}
\end{figure}

The profile publications, as seen in Table \ref{tab1}, contain
inclusions and withdrawals, and also data source issues over which 
there was no control.  So a more suitable analysis, because it was
based on free text search and no more than that, is based on 
the discipline labels.  Therefore our first analysis is of the 
frequency of occurrence data for 16 disciplines crossed by 18 years.  Figure
\ref{figCA} shows the principal factor plane.  Humanities (denoted Hum), 
is off to the left (on the positive side of the ordinate).   

The 
major issue of note in Figure \ref{figCA} is how one-dimensional 
the data is.  In information content expressed by percentage of 
inertia explained by these principal axes, the first axis dominates.  

The (red in the original) lines connect the successive years, that 
are projected onto this principal factor plane.  The clump of years
around Math (mathematics) and Psych (psychology) are all the years
1994 to 2000; and then just north east of them is 2001, followed by 
2002, and north west of it, 2003.  They are all on the negative side
of the first factor, in terms of projections on that factor.  Over 
to the positive side of the first factor are the years 2004 through to 
2011.  The years that lie out a little to the north of the clump 
are 2011 and (the further away year) 2008.  

Can we say that Math and Psych, and Soc (sociology) are more typical 
of the earlier years in regard to cluster analysis research; and that 
if anything Mgt (management) is most typical of the later years here, 
in regard to computer science research?  In order to address this, 
we ought to look at the full dimensionality of the data rather than 
just a 2-dimensional projection.   This will be done below.  

Profile publications are shown as dots in Figure \ref{figCA}
(so as not to crowd the figure).   Some are projected well 
off this figure.  Because these are cited publications rather 
than coming from a given discipline, let us look at them through 
the cluster analyses to follow now.  

%NOTE - medicine slightly repositioned. 
%plot(xc2$cproj[,1], xc2$cproj[,2], type=''n'', xlim=c(-1.0,0.8), ylim=c(-0.2, 0.5),
%+     xlab=''Factor 1, 92% of inertia'', ylab=''Factor 2, 2.6% of inertia'')
%plaxes(c(-1.0,0.8), c(-0.2, 0.5))
%yyy <- xc2$rproj[,2]
%yyy[1] <- yyy[1]- 0.02
%text(xc2$rproj[,1], yyy, dscpl)
%points(xc2$cproj[,1], xc2$cproj[,2], type=''b'', col=''red'')
%text(xc2supp[,1], xc2supp[,2], rep(``.'', 82))
%title(``16 disciplines; 18 years (connected); 82 publications (dots)'')

\subsection{Semantics Analyzed through Clustering of 
Years, Disciplines and Publications}

In the following we use the Euclidean space, with 
equiweighted points, as provided by the Correspondence
Analysis.  The points in this space are equiweighted.  
Furthermore we use the full dimensionality Euclidean space.
Since the active analysis was on the 16 disciplines crossed by 
18 years data, the full dimensionality of the Euclidean factor
space is min $(16 - 1, 18 - 1)$, i.e.\ 15.  In this 15-dimensional
space (illustrated by the planar projection in Figure \ref{figCA}
we thus have disciplines and years projected, and then as passive
(or post hoc) elements we have publications projected.  Because 
the projection takes full account of interrelationships 
as discussed in subsection \ref{semantic} we have that years, 
disciplines and publications are all projected into the same 
space.  
 
Ward's minimum variance hierarchical clustering using Euclidean
distances is an appropriate method to use.  It is appropriate in 
the sense that it uses aggregation based on inertia (masses all
identical) which dovetails with the inertia-based decomposition of
the Correspondence Analysis.  (This hierarchical clustering criterion
was initially described by Joe H. Ward Jr., who died on 23 June 2011,
aged 84.) 

\begin{figure}
\begin{center}
\includegraphics[width=12cm]{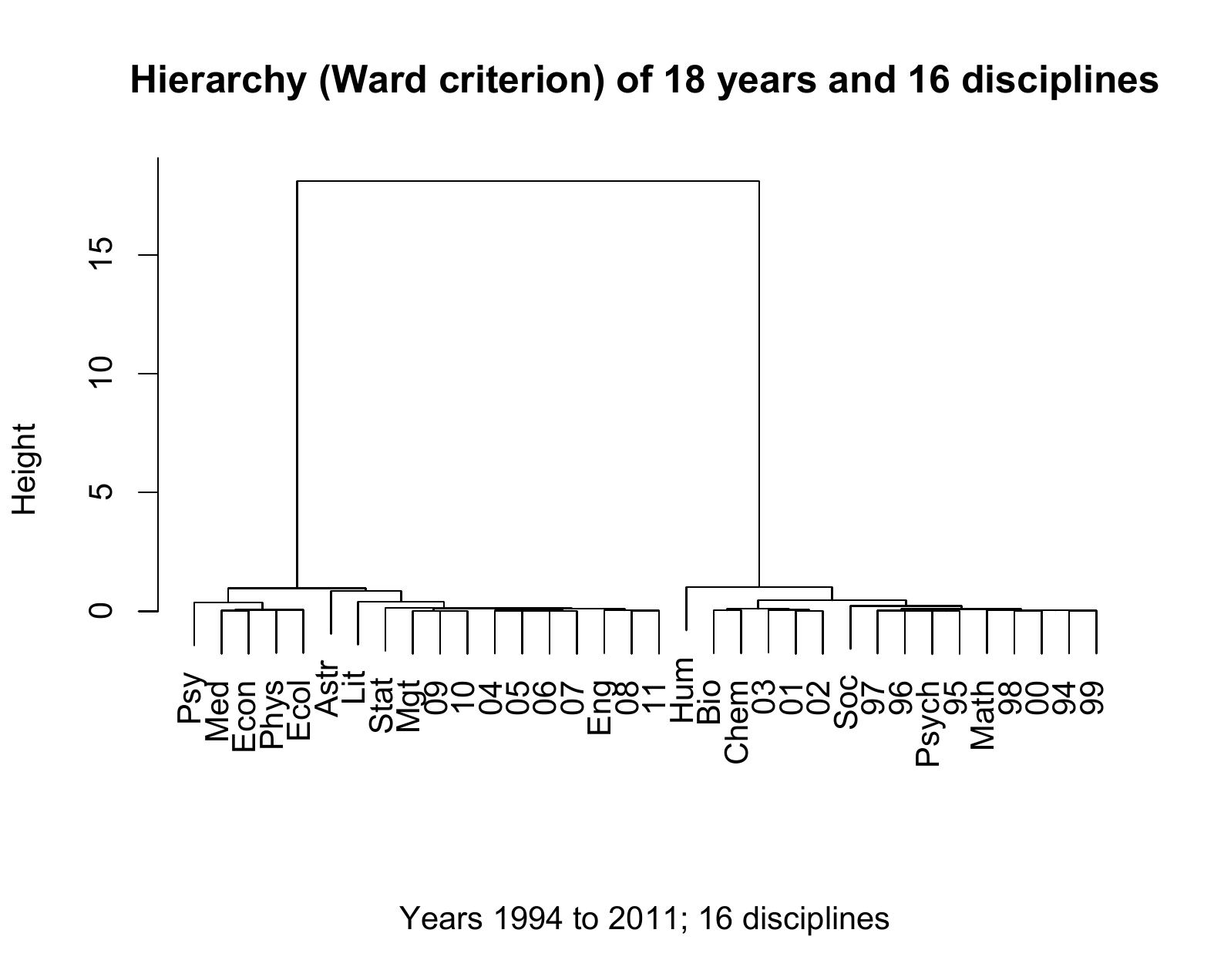}
\end{center}
\caption{Hierarchical clustering of the years and the disciplines. 
(Relative horizontal positioning of terminals is for display purposes only.)}
\label{hc1}
\end{figure}

Figure \ref{hc1} relates to disciplines and years.   The very clear
year-based division of the data is displayed by the two big branches 
in the dendrogram.  We also have further support of the quite 
key role of Psych (psychology) and Math (mathematics), and others,
in the early years; and the key role of Mgt (management), Stat 
(statistics) less pronounced but present, and others, in the later years.  

\begin{figure}
\begin{center}
\includegraphics[width=11cm]{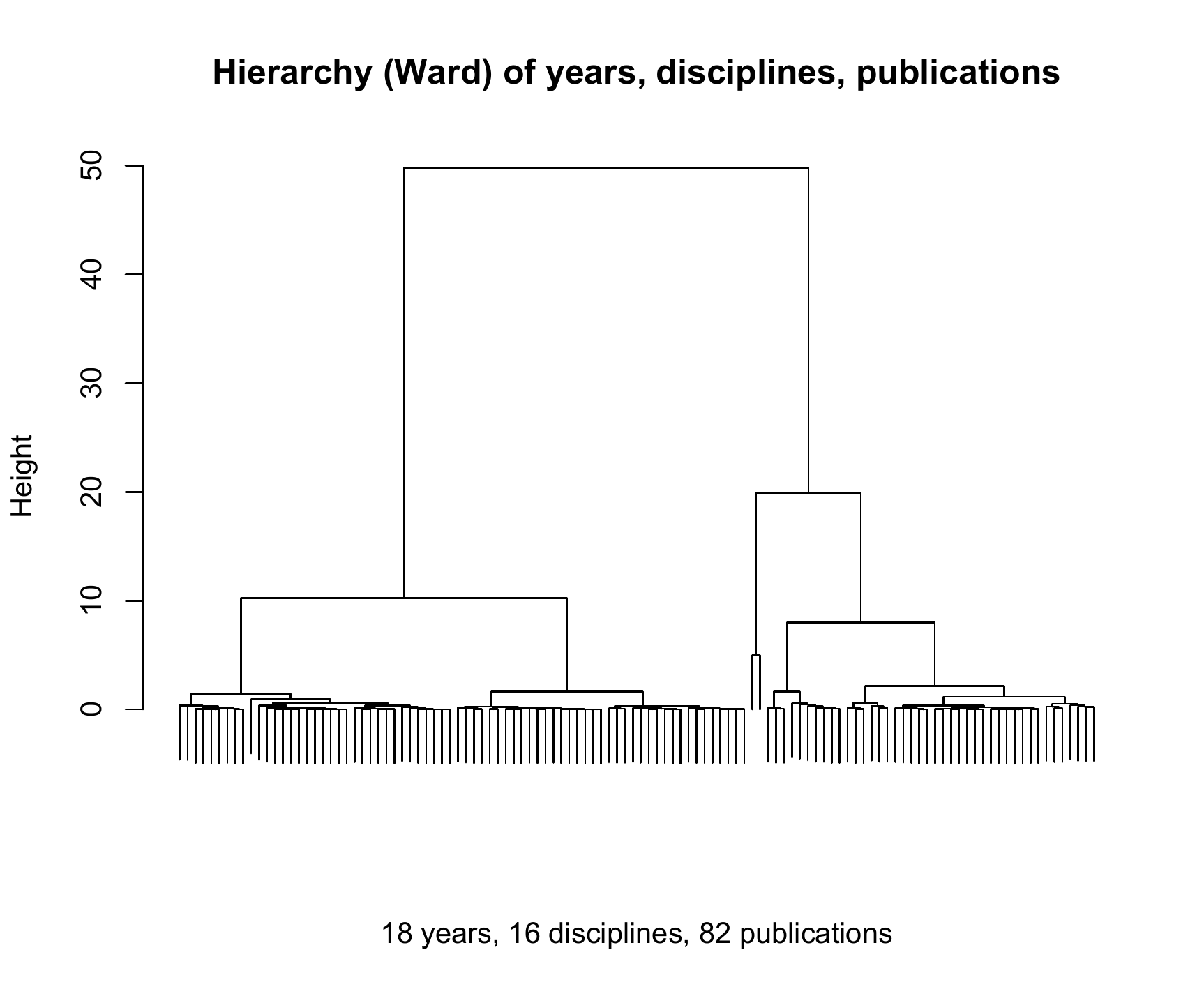}
\end{center}
\caption{Hierarchical clustering of the years, the disciplines, and 
the profile documents.}
\label{hc2}
\end{figure}

\begin{figure}
\begin{center}
\includegraphics[width=11cm]{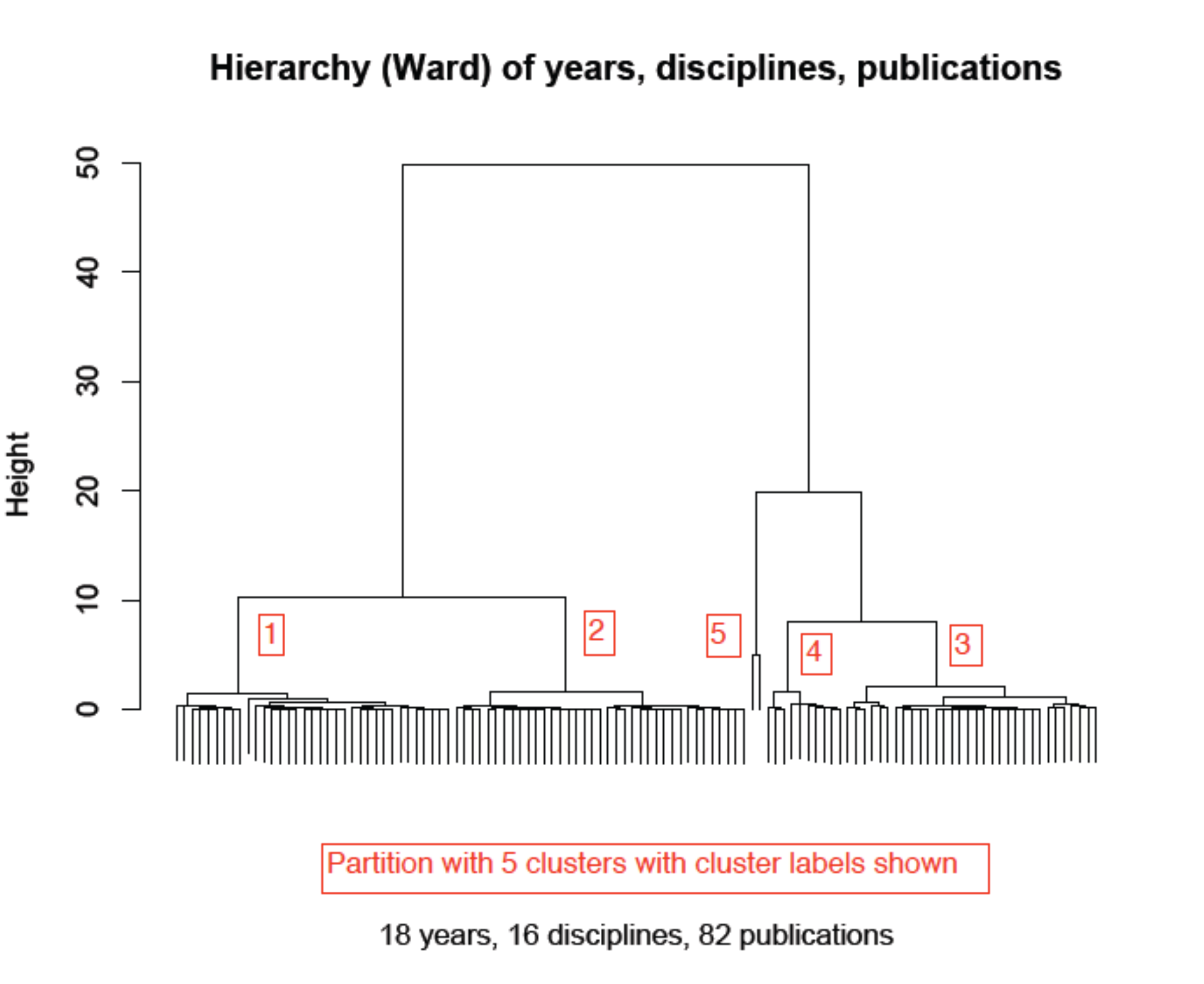}
\end{center}
\caption{As Figure \ref{hc2}, where clusters of the  5-class partition 
are labelled.}
\label{hc2a}
\end{figure}

In Figure \ref{hc2}, the 82 documents are also included.  For 
discussion of clusters, we will use the labels shown in Figure 
\ref{hc2a}.  This allows us more easily to discuss the 
publications, and their associations with years and disciplines, in 
order to home in on major trends and patterns in this data.  

\begin{itemize}
\item[Cluster 1] (cf.\ Figure \ref{hc2a}):

\begin{itemize}
\item[Disciplines:] Med Phys Astr Eng Ecol Psy Lit Econ Mgt
\item[Years:] 04 05 06 07 08 09 10 11
\item[Publications:] 
Bezdek81 Blashfield76 Duda73 Everitt79,80 Fisher36 Fukunaga72 
 Gordon81 Hand81 Hubert7685 Jain88 Kohonen95 Mayr69 McLachlan88,92,97 
 Murtagh83 Pavlidis77 Rand71 Ripley81 Sokal63
\end{itemize}

\item[Cluster 2:]   \ \ 

\begin{itemize}
\item[Disciplines:] Bio Chem Stat 
\item[Publications:] Breiman84 Cormack71 Devijver82 Diggle83 Efron83
 Eldredge80 Friedman77 Fu74,82 Gower66 Hartigan75 Hennig66 Jardine71 
 Johnson67 Kluge69 Kruskal64,78 Lance67 Legendre83 Lorr83 Mantel67 
 Milligan80,81,85 Nei72 Nelson81 Orloci78 Punj83 Reyment84 Sankoff83
 Silverman86 Spaeth80 Tversky77 VanLaarhoven87 Ward63 Wiley81 Wolfe70
 Zahn71
\end{itemize}

\item[Cluster 3:] \ \ 

\begin{itemize}
\item[Disciplines:] Math Psych Soc
\item[Years:] 94 95 96 97 98 99 00 01 02 03
\item[Publications:] Adams72 Anderberg73 Avise74 Benzecri73 Farris72 
 Felsenstein82 Fitch67 Greenacre84 Guttman68 Hill74 Michalski83
 Nosofsky84 Rohlf82 Sattath77 Schiffman81 Sneath73 Spitzer74
 Swofford81 Wishart87
\end{itemize}

\item[Cluster 4:] \ \ 

\begin{itemize}
\item[Discipline:] Hum
\item[Publications:] Arabie87 Carroll70,80 Cover67 Gauch82 Gnanadesikan77
 Huber85 Maddison84 Rammal86 Sammon69
\end{itemize}

\item[Cluster 5:] \ \ 

\begin{itemize}
\item[Publications:] Bishop95 VanRijsbergen79
\end{itemize}

\end{itemize}

From these clusters it can be seen how the ``classical''
period characterized by cluster 3 is counterposed to the
``modern'' period of cluster 1.  

The dominant disciplines of 
the ``classical'' period were Math, Psych and Soc (mathematics,
psychology and sociology).  Certainly some of the profile publications
cited in the ``classical'' period come from ecology, phylogeny and 
even machine learning, but this is not a matter of their 
disciplines but rather cross-discipline influence.   

For the ``modern'' period, cluster 1, it is seen in the planar
projection of Figure \ref{figCA} how Mgt, management, is very 
central.  Other disciplines that characterize especially this 
cluster are noted above.  The more influential profile 
publications can be read off too.

Clusters 4 and 5 are broadly associated with the ``classical'' 
period.  The pattern recognition and information retrieval 
profile publications of cluster 5 are in tune with this 
(given the major ongoing role certainly from the 1960s 
of these sub-disciplines).  

Cluster 2, closest to the ``modern'' period, is characterized
most of all by the disciplines of Bio, Chem, Stat, viz.\ biology,
chemistry and statistics.   See how in
Figure \ref{figCA}, we would not have found that outcome from the 
planar projection alone.  

\section{A Search User Interface}

The open source Apache Solr indexing, querying and search system 
(Solr, 2013a) was used.  Version 4.0 was used in our work. 

\begin{figure}
\begin{center}
\includegraphics[width=16cm]{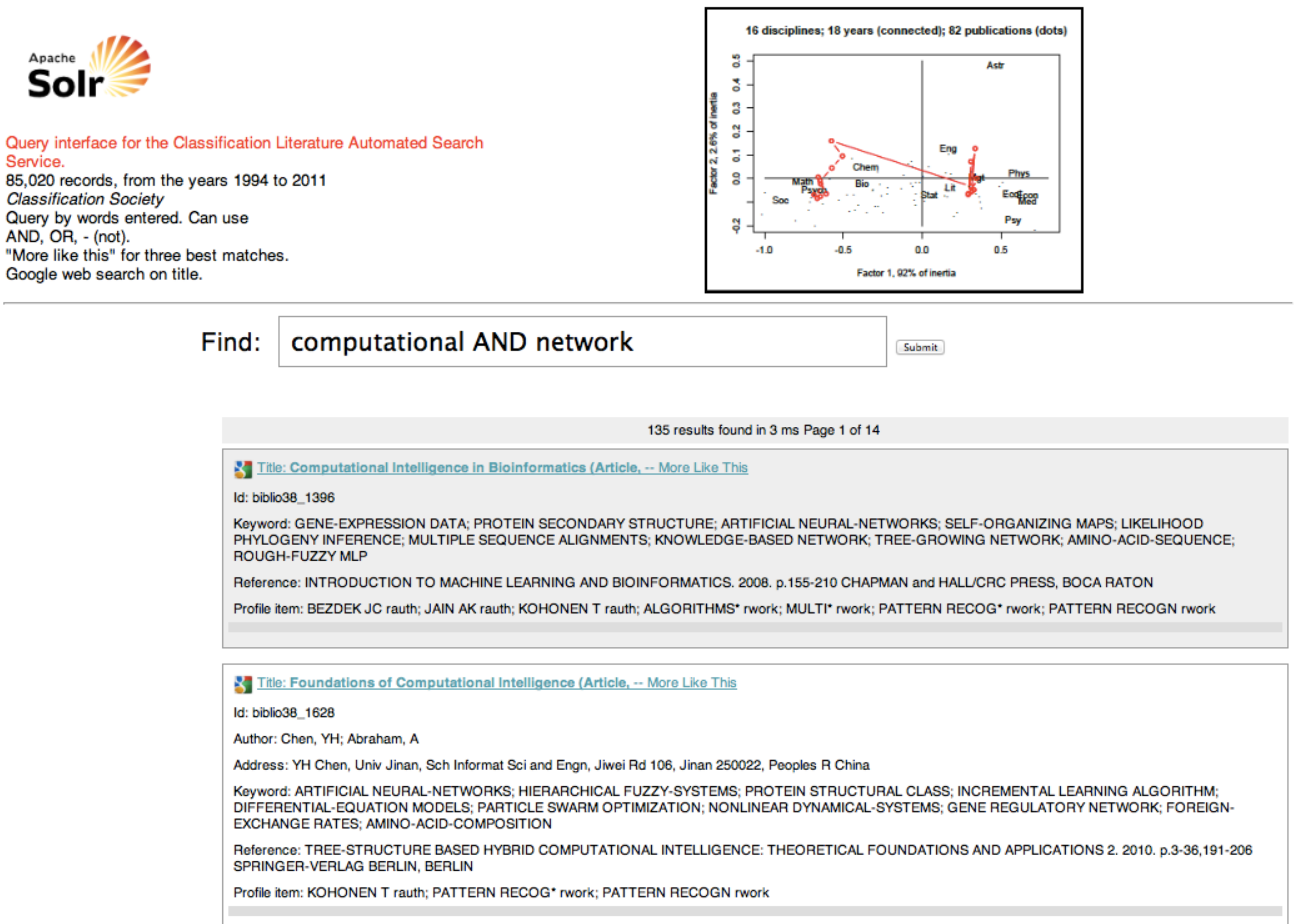}
\end{center}
\caption{Example screenshot of use of Solr search.  The query 
submitted is ``computational AND network'' where these terms can 
appear in any field.  The ``More Like This'' facility provides 
three close matches, on the basis of words used in any field. 
Clicking on the document title provides output from a Google 
search on the web, using all words in the title.}  
\label{figsolr}
\end{figure}
%Used in figure:
%Start server remotely, java -jar start.jar
%http://www.multiresolutions.com:8983/solr/collection1/browse?q=computational+AND+network

Figure \ref{figsolr} shows a screen in a sample session.  All of 
the following are supported: querying repeatedly; 
constraining the query by field (e.g.\ ``author:Arabie''); having a web-wide
Google search carried out in another browser screen for the set
of words appearing in a title; and finding three ``More Like This'' 
results for a given bibliographic record, based on a weighted 
(for the fields) set of common words.  There is practically no 
latency in having query results returned and displayed.   As set up,
10 results are shown per page, and successive records are displayed
with grey and with white background.  

Solr is an enterprise search and display facility, implying that
its secure use needs to be within an enterprise (i.e.\ firewall
or access) setting.  See Solr (2013b).

\section{Conclusions}

The 135,088 citations to one or more of the 82 profile publications
have led us to find a major thematic shift in clustering research over
the 18 years considered here.  At its most basic, this thematic
shift is from the central role of mathematical psychology in the years
1994 to 2003, and then the central role subsequently of management. 
A trend of massive proportions has also been the 
annual increase in {\em Service} contents.  
Other less pronounced trends can be noted also.   

Cluster analysis has shown, and continues to show, great vitality in 
terms of responding to the challenges raised in many different disciplines.   
Vitality is both methodological and practical.   

\section*{Acknowledgement}

Eva Whitmore's role as Technical Editor from the academic year 
1988--1989 to date (mid-2012) has been greatly appreciated.  
%Bill Day's checking of the historical details are also gratefully
%acknowledged.  

\section*{References}
%\begin{thebibliography}{99}

\noindent
%\bibitem{ccs1998}
CCS (1998). 
Computing Classification System, Association for 
Computing Machinery, ACM, 1998.  http://www.acm.org/about/class/ccs98-html
(Viewed: 2013-03-03.)

\smallskip

\noindent
%\bibitem{ccs2012}
CCS (2012). Computing Classification System, Association for 
Computing Machinery, ACM, 2012.  http://dl.acm.org/ccs.cfm
(Viewed: 2013-03-03.)

\smallskip

\noindent
%\bibitem{kurtz}
KURTZ, M.J. (1983). 
``Classification methods: an introductory survey'', 
in {\em Statistical Methods in Astronomy}, European Space 
Agency Special Publication 201, pp.\ 47--58.  

\smallskip

\noindent
%\bibitem{murtaghej}
MURTAGH, F. (2008). 
``Origins of modern data analysis linked to the beginnings
and early development of computer science and information engineering'',
{\em Electronic Journal for History of Probability and Statistics},
4 (2), pp.\ 26.  

\smallskip

\noindent
%\bibitem{murtaghhis}
MURTAGH, F. (2013). 
``History of cluster analysis'', in J. Blasius and M.
Greenacre, Eds., {\em The Visualization and Verbalization of Data},
Chapman and Hall, forthcoming.  

\smallskip

\noindent
%\bibitem{solr}
SOLR (2013a). 
Apache Solr, Open source enterprise search platform, version 4.1.  
http://lucene.apache.org/solr  (Viewed: 2013-03-03.)

\smallskip

\noindent
%\bibitem{solrsec}
SOLR (2013b). 
Solr Wiki, Solr security, 
http://wiki.apache.org/solr/SolrSecurity  (Viewed: 2013-03-03.)

%\bibitem{empty}
%PROBABLY REMOVE THIS SECTION IN VIEW OF THE APPENDIX. 

%%\bibitem{hull}
%%D.L. Hull, {\em Science as a Process: An Evolutionary 
%%Account of the Social and Conceptual Development of Science}, 
%%University of Chicago Press, pp.\ xii+586, 1988.

%\bibitem{kurtz}
%M. Kurtz, 

%\bibitem{leroux1} 
%B. Le Roux, ... 

%\bibitem{leroux2}
%B. Le Roux and H. Rouanet, Geometric Data Analysis: From
%Correspondence Analysis to Structured Data Analysis, 
%Kluwer, 2004. 

%\bibitem{murtagh}
%F. Murtagh, Correspondence Analysis ...
%2005. 

%\bibitem{murleg}
%F. Murtagh and P. Legendre, ``Ward's hierarchical clustering 
%method: clustering criterion and agglomerative algorithm'', 
%submitted, 2012. 

%\bibitem{ward}
%J.H. Ward, ``Hierarchical grouping to optimize an objective
%function'', Journal of the American Statistical Association,
%58, 236--244, 1963.  

%\end{thebibliography}

\section*{Appendix: The Profile Publications Used}

\begin{verbatim}
AUTHOR            JOURNAL/BOOK TITLE    VOL P. YR. 

ADAMS EN          SYST ZOOL             21  390 72
ANDERBERG MR      CLUSTER ANAL APPLICA          73
ARABIE P          3 WAY SCALING CLUSTE          87
AVISE JC          SYST ZOOL             23  465 74
BENZECRI JP       ANAL DONNEES                  73
BEZDEK JC         PATTERN RECOGNITION           81
BISHOP CM         NEURAL NETWORKS PATT          95
BLASHFIELD RK     PSYCHOL B             83  377 76
BREIMAN L         CLASSIFICATION REGRE          84
CARROLL JD        ANN R PSYCH           31  607 80
CARROLL JD        PSYCHOMETRI           35  283 70
CORMACK RM        J ROYAL STA A        134  321 71
COVER TM          IEEE INFO T           13   21 67
DEVIJVER PA       PATTERN RECOGNITION           82
DIGGLE PJ         STATISTICAL ANAL SPA          83
DUDA RO           PATTERN CLASSIFICATI          73
EFRON B           J AM STAT A           78  316 83
ELDREDGE N        PHYLOGENETIC PATTERN          80
EVERITT BS        BIOMETRICS            35  169 79
EVERITT BS        CLUSTER ANAL                  80
FARRIS JS         AM NATURAL           106  646 72
FELSENSTEIN J     Q REV BIOL            57  379 82
FISHER RA         ANN EUGENICS 2         7  179 36
FITCH WM          SCIENCE              155  279 67
FRIEDMAN JH       ACM T MATH             3  209 77
FU KS             SYNTACTIC METHODS PA          74
FU KS             SYNTACTIC PATTERN RE          82
FUKUNAGA K        INTRO STATISTICAL PA          72
GAUCH HG          MULTIVARIATE ANAL CO          82
GNANADESIKAN      METHODS STATISTICAL           77
GORDON AD         CLASSIFICATION                81
GOWER JC          BIOMETRIKA            53  325 66
GREENACRE MJ      THEORY APPLICATION C          84
GUTTMAN L         PSYCHOMETRI           33  469 68
HAND DJ           DISCRIMINATION CLASS          81
HARTIGAN JA       CLUSTERING ALGORITHM          75
HENNIG W          PHYLOGENETIC SYSTEMA          66
HILL MO           APPL STAT             23  340 74
HUBER PJ          ANN STATIST           13  435 85
HUBERT L          BR J MATH S           29  190 76
HUBERT LJ         J CLASSIF              2  193 85
JAIN AK           ALGORITHMS CLUSTERIN          88
JARDINE N         MATH TAXONOMY                 71
JOHNSON SC        PSYCHOMETRI           32  241 67
KLUGE AG          SYST ZOOL             18    1 69
KOHONEN T         SELF ORG MAPS                 95
KRUSKAL JB        MULTIDIMENSIONAL SCA          78
KRUSKAL JB        PSYCHOMETRI           29    1 64
LANCE GN          COMPUTER J             9  373 67
LEGENDRE L        NUMERICAL ECOLOGY             83
LORR M            CLUSTER ANAL SOCIAL           83
MADDISON WP       SYST ZOOL             33   83 84
MANTEL N          CANCER RES            27  209 67
MAYR E            PRINCIPLES SYSTEMATI          69
MCLACHLAN GJ      DISCRIMINANT ANAL ST          92
MCLACHLAN GJ      EM ALGORITHM EXTENSI          97
MCLACHLAN GJ      MIXTURE MODELS INFER          88
MICHALSKI RS      MACHINE LEARNING              83
MILLIGAN GW       MULTIV B R            16  379 81
MILLIGAN GW       PSYCHOMETRI           45  325 80
MILLIGAN GW       PSYCHOMETRI           50  159 85
MURTAGH F         COMPUT J              26  354 83
NEI M             AM NATURAL           106  283 72
NELSON G          SYSTEMATICS BIOGEOGR          81
NOSOFSKY RM       J EXP PSY L           10  104 84
ORLOCI L          MULTIVARIATE ANAL VE          78
PAVLIDIS T        STRUCTURAL PATTERN R          77
PUNJ G            J MARKET RES          20  134 83
RAMMAL R          REV M PHYS            58  765 86
RAND WM           J AM STAT A           66  846 71
REYMENT RA        MULTIVARIATE MORPHOM          84
RIPLEY BD         SPATIAL STATISTICS            81
ROHLF FJ          MATH BIOSCI           59  131 82
SAMMON JW         IEEE COMPUT           18  401 69
SANKOFF D         TIME WARPS STRING ED          83
SATTATH S         PSYCHOMETRI           42  319 77
SCHIFFMAN SS      INTRO MULTIDIMENSION          81
SILVERMAN BW      DENSITY ESTIMATION S          86
SNEATH PHA        NUMERICAL TAXONOMY P          73
SOKAL RR          PRINCIPLES NUMERICAL          63
SPATH H           CLUSTER ANAL ALGORIT          80
SPITZER RL        BRIT J PSYCHI        125  341 74
SWOFFORD DL       J HEREDITY            72  281 81
TVERSKY A         PSYCHOL REV           84  327 77
VANLAARHOVEN      SIMULATED ANNEALING           87
VANRIJSBERGEN     INFORMATION RETRIEVA          79
WARD JH           J AM STAT A           58  236 63
WILEY EO          PHYLOGENETICS                 81
WISHART D         CLUSTAN USER MANUAL           87
WOLFE JH          MULTIV B R             5  329 70
ZAHN CT           IEEE COMPUT           20   68 71
\end{verbatim}

\end{document}